\documentstyle[12pt]{article}
\font\titlefont=cmbx10 scaled \magstep3

\def\lprox{\mathrel{\raise .3ex\hbox{$<$\kern-
.75em\lower1ex\hbox{$\sim$}}}}

\def\gprox{\mathrel{\raise .3ex\hbox{$>$\kern-
.75em\lower1ex\hbox{$\sim$}}}}

\begin{document}

\begin{flushright}
\vspace*{-2cm}
 gr-qc/9510071  \\ TUTP-95-4 \\
Oct. 31, 1995
\vspace*{1.5cm}
\end{flushright}

\begin{center}
{\titlefont QUANTUM FIELD THEORY \\
\vspace*{0.1in}
CONSTRAINS TRAVERSABLE\\
\vspace*{0.1in}
WORMHOLE GEOMETRIES}\\
\vskip .7in
L.H. Ford\\
\vskip .2in
Center for Theoretical Physics, Laboratory for Nuclear Science\\
Massachusetts Institute of Technology, Cambridge, Massachusetts 02139\\
\vskip 0.1in
and
\vskip 0.1in
Institute of Cosmology, Department of Physics and Astronomy\\
Tufts University, Medford, Massachusetts 02155\footnote{Permanent address;
email: lford@pearl.tufts.edu}\\
\vskip 0.3in
and
\vskip 0.3in
Thomas A. Roman\\
\vskip .2in
Institute of Cosmology, Department of Physics and Astronomy\\
Tufts University, Medford, Massachusetts 02155\\
\vskip 0.1in
and
\vskip 0.1in
Department of Physics and Earth Sciences\\
Central Connecticut State University, New Britain, CT 06050\footnote{Permanent
 address; email: roman@ccsu.ctstateu.edu}\\
\end{center}

\newpage
\begin{abstract}
   Recently a bound on negative energy densities in four-dimensional
Minkowski spacetime was derived for a minimally coupled, quantized,
massless, scalar field in an arbitrary quantum state. The bound
has the form of an uncertainty principle-type constraint on the magnitude and
duration of the negative energy density seen by a timelike geodesic observer.
When spacetime is curved and/or has boundaries, we argue that the bound
should hold in regions small
compared to the minimum local characteristic radius of curvature or the
distance to any boundaries, since spacetime can be considered
approximately Minkowski on these scales. We apply the
bound to the stress-energy of static traversable wormhole spacetimes.
Our analysis implies that either the wormhole must be only a little larger
than Planck size or that there is a large discrepancy in the length scales
which characterize the wormhole. In the latter case, the negative energy
must typically be concentrated in a thin band many orders of magnitude smaller
than the throat size. These results would seem to make the existence of
macroscopic traversable wormholes very improbable.

\end{abstract}

\newpage
\baselineskip=14pt
\section{Introduction}
\label{sec:intro}

     In recent years there has been considerable interest in the topic
of traversable wormholes, solutions of Einstein's equations which act as
tunnels from one region of spacetime to another, through which an observer
might freely pass \cite{MT,MTY,VBOOK}. Traversable wormhole spacetimes have
the property
that they must involve ``exotic matter'', that is, a stress tensor which
violates the weak energy condition. Thus the energy density must be negative
in the frame of reference of at least some observers. Although classical
forms of matter obey the weak energy condition, it is well-known that quantum
fields can generate locally negative energy densities, which may be arbitrarily
large at a given point. A key issue in the
study of wormholes is the nature and magnitude of the violations of the weak
energy condition which are allowed by quantum field theory. One possible
constraint upon such violations is given by averaged energy conditions
\cite{T78}.
In particular, the averaged null energy condition (ANEC) states that
$\int T_{\mu \nu} k^\mu k^\nu {\rm d}\lambda \geq 0$, where the integral is
taken along a complete null geodesic with tangent vector $k^\mu$ and affine
parameter $\lambda$. This condition must be violated in wormhole spacetimes
\cite{MTY}. Although ANEC can be proven to hold in Minkowski spacetime,
it is generally violated in curved spacetime \cite{WY,Visser}. The extent to
which it can be violated is not yet well understood, but limits on the extent
of ANEC violation will place constraints upon allowable wormhole
geometries \cite{Yurtsever95,FWUNPUB}.

A second type of constraint upon violations of the weak energy condition
are ``quantum inequalities'' (QI's), which limit the magnitude and spatial or
temporal extent of negative energy \cite{F78}-\cite{FR95}. These constraints
are intermediate between pointwise conditions and the averaged energy
conditions
in that they give information about the distribution of negative energy in
a finite neighborhood. For the most part, inequalities of this type have only
been proven flat spacetime. The main purpose of this paper will be to argue
that restricted versions of the flat spacetime inequalities can be
employed in curved spacetime, and that these inequalities place severe
constraints upon wormhole geometries. We assume that the stress-energy of
the wormhole spacetime is a renormalized expectation value of the
energy-momentum tensor operator in some quantum state, and ignore fluctuations
in this expectation value \cite{Kuo,KF}.

In this paper, we restrict our attention to static, spherically symmetric
wormholes. We will also assume that the spacetime contains no closed timelike
curves. This latter assumption may not be necessary, but we make it in order
to insure that quantum field theory on the wormhole spacetime is well-defined.
In Sec.~\ref{sec:QI}, a flat spacetime quantum inequality is reviewed, and
an argument is presented for the application of this inequality in small
regions
of a curved spacetime. In Sec.~\ref{sec:REVMT}, we briefly review some of the
essential features of traversable (Morris-Thorne) wormholes. We next consider
a number of particular wormhole models in Sec.~\ref{sec:SE}, and argue that
the quantum inequality places strong restrictions upon the dimensions of these
wormholes. In Sec.~\ref{sec:GB} we formulate a more general bound upon the
relative dimensions of an arbitrary Morris-Thorne wormhole. Finally, in
Sec.~\ref{sec:con} we summarize and interpret our results.
Our units are taken to be those in which $\hbar=G=c=1$, and our sign
conventions
are those of Ref. \cite{MT}.

\section{Quantum Inequalities in Flat and Curved Spacetime}
\label{sec:QI}

    In Ref. \cite{FR95}, an inequality was proven which limits the magnitude
and duration of the negative energy density seen by an inertial observer in
Minkowski spacetime (without boundaries). Let $\langle T_{\mu\nu} \rangle$
be the renormalized expectation value of the stress tensor for a free,
massless, minimally coupled scalar field in
an arbitrary quantum state. Let $u^\mu$ be the observer's four-velocity, so
that $\langle T_{\mu\nu} u^{\mu} u^{\nu}\rangle$ is the expectation value of
the local energy density in this observer's frame of reference. The inequality
states that
\begin{equation}
{{\tau_0} \over \pi}\, \int_{-\infty}^{\infty}\,
{{\langle T_{\mu\nu} u^{\mu} u^{\nu}\rangle\, d\tau}
\over {{\tau}^2+{\tau_0}^2}} \geq
-{3\over {32 {\pi}^2 {\tau_0}^4}}\,,  \label{eq:QI}
\end{equation}
for all $\tau_0$, where $\tau$ is the observer's proper time.
The Lorentzian function which appears in the integrand is a convenient choice
for a sampling function, which samples the energy density in an interval of
characteristic duration $\tau_0$ centered around an arbitrary point on the
observer's worldline. The proper time coordinate has been chosen so that
this point is at $\tau = 0$.
The physical content of Eq.~(\ref{eq:QI}) is that the more negative the
energy density is in an interval, the shorter must be the duration of the
interval. Consider, for example, a pocket of negative energy which our
observer traverses in a proper time $\Delta \tau$. A natural choice of the
sampling time is $\tau_0 =\Delta \tau$, in which case we infer that the average
value of the negative energy in this pocket is bounded below by
$-3/[32\pi^2(\Delta \tau)^4]$. Because Eq.~(\ref{eq:QI}) holds for all
$\tau_0$, we must obtain a true statement with other choices. If we let
$\tau_0 < \Delta \tau$, then we obtain a weaker bound. If we let
$\tau_0 > \Delta \tau$, then we appear to obtain a stronger bound. However,
now the range over which we are sampling extends beyond the boundaries of
the pocket and may include positive energy contributions. Hence it is to
be expected that the lower bound on the average energy density should be
less negative.

The basic premise of this paper is that one may obtain a constraint upon the
renormalized stress tensor in a curved spacetime using Eq.~(\ref{eq:QI}),
provided that $\tau_0$ is taken to be sufficiently small. The main purpose of
this section is to explore the rationale for this premise. The basic idea is
that a curved spacetime appears flat if restricted to a sufficiently small
region. However, this idea is sufficiently subtle to require an extended
discussion.

First, let us recall the situation in classical general relativity. The
principle of equivalence has its mathematical expression in the fact that
the geodesic equations involve the spacetime metric and the connection
coefficients, but not the curvature tensor. Thus, if we go to a local inertial
frame, the equations of motion for a point test particle take the flat space
form. However, it is possible for equations of motion to contain curvature
terms explicitly. An example is the equation of motion for a classical spinning
test particle \cite{Papa}. In this case, the principle of equivalence does not
hold in its simplest form, and one can treat the system as being in locally
flat spacetime only to the extent that the curvature terms are negligible.

   In quantum field theory, we will be more interested in the extent to
which solutions of wave equations can be approximated by the flat space forms.
Consider, for example, the generalized Klein-Gordon equation
\begin{equation}
\Box \phi + m^2\phi + \xi R \phi =0\,,                  \label{eq:wave}
\end{equation}
where $\xi$ is an arbitrary constant and $R$ is the scalar curvature.
The solutions of this equation will generally not be similar to the flat space
solutions unless the curvature term is small compared to the other terms in
the equation. However, this is still not sufficient to guarantee that a flat
space mode is a solution of Eq.~(\ref{eq:wave}). It is also necessary to
require that the modes have a wavelength that is small compared to the local
radii of curvature of the spacetime. In this limit, it is possible to obtain
WKB-type solutions to Eq.~(\ref{eq:wave}), which are approximately plane wave
modes. For an illustration of this, see the work of Parker and Fulling
\cite{PF}
on adiabatic regularization. These authors give generalized WKB solutions of
wave equations in an expanding spatially flat Robertson-Walker universe.
In the limit that the wavelength of a mode is short compared to the expansion
time scale (which is the spacetime radius of curvature in this case), the
leading term, which is of the plane wave form, becomes a good approximation.

  Our primary concern is when we may expect the inequality Eq.~(\ref{eq:QI}),
which was derived from Minkowski space quantum field theory, to hold in
a curved spacetime and/or one with boundaries. For a given $\tau_0$,
the dominant contribution to the
right-hand side of this inequality arises from modes for which $\lambda \sim
\tau_0$. In particular, modes for which $\lambda \gg \tau_0$ yield a small
contribution. To see this more explicitly, note that the right-hand side
of Eq.~(\ref{eq:QI}) arises from the integral $(4\pi^2)^{-1}\int_0^\infty
{\rm d}\omega\, \omega^3\, {\rm e}^{-2\omega \tau_0}$. (See Eq. (63) of
Ref. \cite{FR95}). Thus if the long wavelength modes ($\omega \ll \tau_0^{-1}$)
were to be omitted or to be distorted by the presence of spacetime curvature
or boundaries, the result would not
change significantly. This suggests that we can apply the inequality in a
curved spacetime so long as $\tau_0$ is restricted to be small compared to
the local proper radii of curvature and the proper distance to any boundaries
in the spacetime. This is the criterion that the
relevant modes be approximated by plane wave modes.

   The specific example of the Casimir effect may be useful as an illustration.
Here one has a constant negative energy density, which would not be possible if
Eq.~(\ref{eq:QI}) holds for all $\tau_0$. However, if we impose some
restrictions
on the allowable values of $\tau_0$, then the inequality {\it does}
in fact still
apply. Let us consider a massless scalar field with periodicity of length $L$
in the $z$-direction. Let us also consider an observer moving with velocity $v$
in the $+z$-direction. In the rest frame of this observer, the expectation
value of the energy density is
\begin{equation}
\langle T_{\mu\nu} u^{\mu} u^{\nu}\rangle = - \frac{\pi^2}{45 L^4}\,
        (1+3v^2)\, \gamma^2 \, ,              \label{eq:enden}
\end{equation}
where $\gamma = (1-v^2)^{-\frac{1}{2}}$. Because this quantity is a constant,
Eq.~(\ref{eq:QI}) becomes
\begin{equation}
- \frac{\pi^2}{45 L^4}\, (1+3v^2)\, \gamma^2 \geq
      -{3\over {32 {\pi}^2 {\tau_0}^4}}\,,
\end{equation}
or equivalently,
\begin{equation}
\tau_0 \leq \frac{3L}{2\pi}\, \biggl(\frac{5}{6}\biggr)^{\frac{1}{4}}\,
 [(1+3v^2)\, \gamma^2]^{-\frac{1}{4}} \, .   \label{eq:tau0}
\end{equation}
Thus for the special case of a static observer ($v=0$), we must have
\begin{equation}
\tau_0 \leq \frac{3L}{2\pi}\, \biggl(\frac{5}{6}\biggr)^{\frac{1}{4}}
                              \approx 0.46\,L\, .     \label{eq:tau0'}
\end{equation}

  There are two relevant length scales in the observer's frame of reference.
The first is the (Lorentz-contracted) periodicity length, $l_1 = L/\gamma$,
and the second is the proper time required to traverse this distance,
$l_2 = L/(v\gamma)$. Here $l_1$ is the smaller of the two, and plays a
role analogous to the minimum radius of curvature in a curved spacetime.
Thus we should let
\begin{equation}
\tau_0 = f l_1 = \frac{fL}{\gamma}\,.
\end{equation}
Equation~(\ref{eq:tau0}) will be satisfied if
\begin{equation}
f \leq g(v) \equiv \frac{3}{2\pi}\, \biggl(\frac{5}{6}\biggr)^{\frac{1}{4}}\,
 [(1+3v^2)(1-v^2)]^{-\frac{1}{4}} \, .   \label{eq:f}
\end{equation}
The function  $g(v)$ has its minimum value at $v = 1/\sqrt{3}$, at which
point
\begin{equation}
g\biggl(\frac{1}{\sqrt{3}}\biggr) =
\frac{3}{2\pi}\, \biggl(\frac{5}{8}\biggr)^{\frac{1}{4}} \approx 0.42 \,.
                                           \label{eq:tau0''}
\end{equation}
Thus if we restrict $\tau_0 < 0.42\, l_1$, then the Minkowski space quantum
inequality also holds in the compactified spacetime. Note that the constraint
obtained by considering arbitrary $v$ differs only slightly from that for
static observers, Eq.~(\ref{eq:tau0'}).

   The Casimir effect example contains some of the essential features that
we encounter in a renormalized stress tensor on a curved background spacetime.
However, on a curved spacetime $\langle T_{\mu\nu} \rangle$ is a sum of a
state-dependent part and a state-independent geometrical part. The latter
consists of terms which are
either quadratic in the Riemann tensor or else linear in second derivatives
of the Riemann tensor. One source of curvature dependence in
$\langle T_{\mu\nu} \rangle$ is the well-known trace anomaly.
For the case of the conformal ($\xi =1/6$) scalar field, it is
\begin{equation}
\langle T^\mu_\mu \rangle = {1\over {2880\pi^2}} \Bigl(
R_{\alpha\beta\rho\sigma}R^{\alpha\beta\rho\sigma}
- R_{\alpha\beta}R^{\alpha\beta} + \nabla_\rho \nabla^\rho R \Bigr).
                                            \label{eq:trace}
\end{equation}
Other fields have trace anomalies with similar coefficients,  i.e.,
with magnitudes of the order of $10^{-4}$.  Thus these terms will give a
very small contribution to a quantum inequality of the form of
Eq.~(\ref{eq:QI})
when $\tau_0 \ll l$, where $l$ is the characteristic radius of curvature.

   A related source of curvature dependence in the renormalized stress tensor
is the possible presence of finite terms of the form of the quadratic
counter-terms required to remove the logarithmic divergences in a curved
spacetime. These terms are the tensors
\begin{eqnarray}
H^{(1)}_{\mu\nu} &\equiv& {1\over \sqrt{-g}} {\delta \over {\delta g^{\mu\nu}}}
                                 \bigl[\sqrt{-g} R^2 \bigr] \nonumber \\
&=& 2\nabla_\nu \nabla_\mu R -2g_{\mu\nu}\nabla_\rho \nabla^\rho R
 + {1\over 2}g_{\mu\nu} R^2 - 2R R_{\mu\nu},  \label{eq:H1}
\end{eqnarray}
and
\begin{eqnarray}
H^{(2)}_{\mu\nu} &\equiv&
                {1\over \sqrt{-g}} {\delta \over {\delta g^{\mu\nu}}}
            \bigl[\sqrt{-g} R_{\alpha\beta}R^{\alpha\beta} \bigr]
= 2\nabla_\alpha \nabla_\nu R_\mu^\alpha +\nabla_\rho \nabla^\rho R_{\mu\nu}
\nonumber \\ &{}& -{1\over 2}g_{\mu\nu}\nabla_\rho \nabla^\rho R
  +{1\over 2}g_{\mu\nu} R_{\alpha\beta}R^{\alpha\beta}
   - 2R_\mu^\rho R_{\rho\nu}.    \label{eq:H2}
\end{eqnarray}
There can be a term of the form $c_1 H^{(1)}_{\mu\nu} + c_2 H^{(2)}_{\mu\nu}$
in $\langle T_{\mu\nu} \rangle$. More generally, there might be a term of the
form $(c_1 H^{(1)}_{\mu\nu} + c_2 H^{(2)}_{\mu\nu}) {\rm log}(R \mu^{-2})$,
where $\mu$ is an arbitrary renormalization mass scale \cite{FT}.
A shift in the value of
$\mu$ adds a term proportional to $c_1 H^{(1)}_{\mu\nu} + c_2 H^{(2)}_{\mu\nu}$
to $\langle T_{\mu\nu} \rangle$. Visser \cite{Visser} has recently discussed
how terms of this form are likely to lead to violations of ANEC in curved
spacetime. The problem is that quantum field theory by itself is not able to
predict the values of $c_1$ and $c_2$, or equivalently, of $\mu$. Thus very
large values of these parameters are not, {\it a priori}, ruled out.
The status of these terms in the semiclassical Einstein equations has been the
subject of much discussion in the literature. They appear to give rise to
unstable behavior \cite{Horowitz}, analogous to the runaway solutions of the
Lorentz-Dirac equation of classical electron theory. More recently Simon
\cite{Simon} has suggested that it may be possible to reformulate the
semiclassical theory to avoid unstable solutions.

If one ignores the possibility of runaway solutions, then
if these terms are to produce a significant correction to the geometry of a
spacetime whose curvature is far below Planck dimensions, then at least one
of the dimensionless constants $c_1$ or $c_2$ must be extremely large.
The Einstein tensor is of order $l^{-2}$ and the $H^{(1)}_{\mu\nu}$
and $H^{(2)}_{\mu\nu}$ tensors are of order $l^{-4}$, in Planck units.
The latter are negligible
unless their coefficients are at least of order $(l/l_p)^2$,
where $l_p$ is the Planck length. Thus if the state-independent geometrical
part of
$\langle T_{\mu\nu} \rangle$ is to be the source of the exotic matter
which generates the wormhole geometry, either the wormhole must be of
Planck dimensions, or else one must accept large dimensionless coefficients.
For example,
unless one of these constants is at least of order $10^{70}$, the quadratic
curvature terms will be negligible for the discussion of a wormhole whose
throat radius is of the order of $1\, m$. A value of $c_1$ or $c_2$ of
$10^{70}$
could arise from a single quantum field or from $10^{70}$ fields each giving
a contribution of order unity \cite{LOG}. Both possibilities seem equally
unnatural.

An alternative is for the state-dependent part of $\langle T_{\mu\nu} \rangle$
to be the source of the exotic matter. A non-exotic stress tensor may be made
arbitrarily large by increasing the particle content of the quantum state.
One might naively expect that the same could be done for a stress tensor
representing exotic matter. However, the essential content of the quantum
inequality Eq.~(\ref{eq:QI}) is that arbitrarily extended distributions
of arbitrarily negative energy are not possible in Minkowski
spacetime. In this section we have argued that the bound should also be
applicable in curved spacetimes for sampling times small compared
to either the minimum local radius of curvature or the proper distance to
any boundary.

Let us recall that Eq.~(\ref{eq:QI}) was proven for the specific case of a
free massless, minimally coupled scalar field. It should be straightforward
to generalize the
arguments of Ref. \cite{FR95} to the case of other massless fields, such as
the electromagnetic field. Although this has not yet been done, it is unlikely
that the result will be significantly different. Generalizations to massive
fields may also be possible, although the results may be more complicated
due to the presence of  two length scales, $\tau_0$ and the particle's Compton
wavelength. However, it
seems unlikely that adding a mass will make it easier to have large negative
energy densities, as one now has to overcome the positive rest mass energy.
Thus, one suspects that massive fields will satisfy inequalities which are
more restrictive than Eq.~(\ref{eq:QI}). The effect of including interactions
is the most difficult to assess. If an interacting theory were to allow regions
of negative energy much more extensive than allowed in free theories, there
would seem to be a danger of an instability where the system spontaneously
makes a transition to configuration with large negative energy density.
However,
this must be regarded as an open question.

\section{Morris-Thorne Wormholes}
\label{sec:REVMT}

 The spacetime geometry for an MT traversable wormhole is described by
the metric \cite{MT}:
\begin{equation}
 ds^2=-e^{2\Phi(r)}dt^2+{{dr^2}\over{(1-b(r)/r)}}
           +r^2({d\theta}^2+ {\rm sin}^2\theta\,{d\phi}^2)\,,
                                                \label{eq:MTM1}
\end{equation}
where the two adjustable functions $b(r)$ and $\Phi(r)$ are
the ``shape function'' and the ``redshift
function'', respectively. The shape function $b(r)$ determines the
shape of the wormhole as viewed, for example, in an embedding
diagram. The metric Eq.~(\ref{eq:MTM1}) is spherically symmetric and static,
with the proper circumference of a circle of fixed $r$ being given by
$2\pi r$. The coordinate $r$
is non-monotonic in that it decreases from $+\infty$ to a minimum
value $r_0$, representing the location of the throat of the
wormhole, where $b(r_0)=r_0$, and then it increases from $r_0$ to $+\infty$.
Although there is a coordinate singularity at the throat, where the
metric coefficient $g_{rr}$ becomes divergent, the radial
proper distance
\begin{equation}
 l(r)=\pm\,\int_{r_0}^r{{dr}\over{(1-b(r)/r)^{1/2}}} \,,
                                                 \label{eq:PD}
\end{equation}
is required to be finite everywhere. Note that because
$0 \leq 1 -b(r)/r \leq 1$, the proper distance is greater than or equal to
the coordinate distance: $|l(r)| \geq r - r_0$.   The metric
Eq.~(\ref{eq:MTM1}) may be written in terms of the proper radial distance as:
\begin{equation}
ds^2=-e^{2\Phi(r)}dt^2 + dl^2
                + {r^2}(l)({d\theta}^2+ {\rm sin}^2\theta\,{d\phi}^2)\,.
                                                \label{eq:MTM2}
\end{equation}
The proper distance decreases
from $l=+\infty$ to zero at the throat, and then from zero to
$-\infty$ on the ``other side'' of the wormhole. For the wormhole to
be traversable it must have
no horizons, which implies that $g_{tt}=-e^{2\Phi(r)}$ must never
be allowed to vanish, and hence $\Phi(r)$ must be everywhere finite.

The four-velocity of a static observer is $u^{\mu}=dx^{\mu}/{d\tau}
=(u^{\,t},0,0,0)=(e^{-\Phi(r)},0,0,0)$. The observer's four-acceleration is
\begin{eqnarray}
a^{\mu} &=& {{Du^{\mu}}\over {d\tau}}  \nonumber  \\
        &=& u^{\mu}\,_{;\,\nu}\,u^{\nu}  \nonumber \\
 &=& (u^{\mu}\,_{,\,\nu}+\Gamma^{\mu}_{\beta\nu}\,u^{\beta})\,u^{\nu} \,.
                                                   \label{eq:4acc}
\end{eqnarray}
For the metric Eq.~(\ref{eq:MTM1}) we have
\begin{eqnarray}
a^t &=& 0 \,,         \nonumber \\
a^r &=& \Gamma^r_{tt}\,\left({dt\over{d\tau}}\right)^2
    ={\Phi}'\,(1-b/r)\,,          \label{eq:4acc-comp}
\end{eqnarray}
where $\Phi' = d\Phi/dr$.
{}From the geodesic equation, a radially moving test particle which
starts from rest initially has the equation of motion
\begin{equation}
{{d^{\,2}r}\over d{\tau}^2}=-\Gamma^r_{tt}\,
\left({dt\over{d\tau}}\right)^2 =-a^r\,.
\end{equation}
Hence $a^r$ is the radial component of proper
acceleration that an observer must maintain in order to remain at
rest at constant $r,\,\theta,\,\phi$. Note for future reference that from
Eq.~(\ref{eq:4acc-comp}), {\it a static observer at the throat of any
wormhole is a geodesic observer}. For $\Phi'(r)\neq 0$ wormholes,
static observers are not geodesic (except at the throat),
whereas for $\Phi'(r)=0$ wormholes they are. A wormhole is
``attractive'' if $a^r>0$ (observers must maintain an outward-directed
radial acceleration to keep from being pulled into the wormhole),
and ``repulsive'' if $a^r<0$ (observers must maintain an inward-directed
radial acceleration to avoid being pushed away from the wormhole). From
Eq.~(\ref{eq:4acc-comp}), this distinction depends on the sign of $\Phi'$.
For $a^r=0$, the wormhole is neither attractive nor repulsive.

Substitution of Eq.~(\ref{eq:MTM1}) into the Einstein equations gives
the stress-energy tensor required to generate the wormhole geometry. It
is often convenient to work in the static orthonormal frame given by the
basis:
\begin{eqnarray}
    e_{\hat{t}} &=& e^{-\Phi}\,e_{t}, \nonumber \\
    e_{\hat{r}} &=& (1-b/r)^{1/2}\,e_{r}, \nonumber \\
    e_{\hat{\theta}} &=& r^{-1}\,e_{\theta}, \nonumber \\
    e_{\hat{\phi}} &=& (r\,{\rm sin}\theta)^{-1}\,e_{\phi}\,.
                                                   \label{eq:SOBASIS}
\end{eqnarray}
This basis represents the proper reference frame of an observer who
is at rest relative to the wormhole. In this frame the stress tensor
components are given by
\begin{eqnarray}
T_{\hat t \hat t} &=& \rho
 ={b' \over {8\pi r^2} } \,, \label{eq:Ttt} \\
 T_{\hat r \hat r} &=& p_r
 =-{1 \over {8\pi} } \, \biggl[{ b \over {r^3} }-{ {2{\Phi}'} \over r }\,
 \biggl(1-{b \over r} \biggr) \biggr] \,, \label{eq:Trr} \\
 T_{\hat \theta \hat \theta} &=& T_{\hat \phi \hat \phi} = P \nonumber \\
&=&\!\!\!\!{1 \over {8\pi} } \,\biggl[{1 \over 2 } \biggl(
{ b \over {r^3} } - { {b'} \over {r^2} } \biggr) +
{ { {\Phi}'} \over r } \,
 \biggl(1- { b \over {2r} } - { b'\over 2 } \biggr) +
  \biggl(1-{b \over r} \biggr) \, ({\Phi}''+({\Phi}')^2) \biggr] \,.
                                        \label{eq:Tang}
\end{eqnarray}
The quantities $\rho,\, p_r$, and $P$ are the
mass-energy density, radial pressure, and transverse
pressure, respectively, as measured by a static observer \cite{MC}.
At the throat of the wormhole, $r=r_0$, these reduce to
\begin{eqnarray}
{\rho}_0 &=& {{b'_0}\over{8\pi {r_0}^2}}  \,,  \label{rho;r0} \\
p_0 &=& - {1 \over{8\pi {r_0}^2}} \,,      \label{p_r;r0} \\
P_0 &=& {{1 - b'_0} \over {16\pi r_0}} \,
\biggl(\Phi'_0 + {1 \over {r_0} } \biggr) \,,  \label{P;r0}
\end{eqnarray}
where $b'_0 = b'(r_0)$ and $\Phi'_0 = \Phi'(r_0)$.

The curvature tensor components are given by
\begin{eqnarray}
R_{\hat{t}\hat{r}\hat{t}\hat{r}}
&=& \biggl(1- {b \over r} \biggr)\,[{\Phi}''\,+\,{({\Phi}')}^2]\,
+\, { {\Phi}' \over {2r^2} } \, (b - b'r)  \,,    \label{eq:Rtrtr}  \\
R_{\hat{t}\hat{\theta}\hat{t}\hat{\theta}}
&=& R_{\hat{t}\hat{\phi}\hat{t}\hat{\phi}}
= { {\Phi}'\over r}\,\biggl(1- {b \over r} \biggr)  \,,
                                            \label{eq:Rttheta} \\
R_{\hat{r}\hat{\theta}\hat{r}\hat{\theta}}
&=& R_{\hat{r}\hat{\phi}\hat{r}\hat{\phi}}
= {1 \over {2 r^3} } \, (b'r-b) \,,   \label{eq:Rrtheta}  \\
R_{\hat{\theta}\hat{\phi}\hat{\theta}\hat{\phi}}
&=& {b \over {r^3} } \,.             \label{eq:Rthetaphi}
\end{eqnarray}
All other components of the curvature tensor vanish, except for those
related to the above by symmetry. At the throat, these
components reduce to
\begin{eqnarray}
R_{\hat{t}\hat{r}\hat{t}\hat{r}}|_{r_0}
&=& { {\Phi'_0} \over {2r_0} } \, (1 - b'_0)  \,,    \label{eq:Rtrtr|r_0}  \\
R_{\hat{t}\hat{\theta}\hat{t}\hat{\theta}}|_{r_0}
&=& R_{\hat{t}\hat{\phi}\hat{t}\hat{\phi}}|_{r_0}
= 0 \,,                           \label{eq:Rttheta|r_0} \\
R_{\hat{r}\hat{\theta}\hat{r}\hat{\theta}}|_{r_0}
&=& R_{\hat{r}\hat{\phi}\hat{r}\hat{\phi}}|_{r_0}
= -{1 \over {2 {r_0}^2} } \, (1 - b'_0) \,,   \label{eq:Rrtheta|r_0}  \\
R_{\hat{\theta}\hat{\phi}\hat{\theta}\hat{\phi}}|_{r_0}
&=& {1 \over {{r_0}^2} } \,.             \label{eq:Rthetaphi|r_0}
\end{eqnarray}

Let us now define the following set of length scales:
\begin{equation}
{\bar r_0} = b  \,;\,\, r_1 = \biggl|{b \over {b'} }\biggr|  \,;\,\,
r_2 = \biggl|{ {\Phi} \over {\Phi'} }\biggr|  \,;\,\,
r_3 = \biggl|{ {\Phi'} \over {\Phi''} }\biggr| \,.   \label{eq:ls}
\end{equation}
The quantities $r_1,r_2,r_3$ are a measure of the coordinate length scales
over which
$b,\Phi$, and $\Phi'$, respectively, change. The number of length scales
correspond to the number of derivatives which appear in the curvature
tensor, and $b$. It will prove convenient to absorb $|\Phi|$ into
another length scale defined by
\begin{equation}
R_2 = { {r_2} \over {|\Phi|} } = { 1 \over {|\Phi'|} } \,. \label{eq:R_2}
\end{equation}
The smallest of the above length scales is
\begin{equation}
r_m \equiv min({\bar r_0}, r_1, R_2, r_3) \,.  \label{eq:rmin}
\end{equation}
As an aside, note that if $r_m=R_2$, then we can say that either
$r_2$ is very small or $|\Phi|$ is very large (which implies that
the redshift/blueshift, $e^{\, \pm |\Phi|}$, is very large), or both.
The curvature components may be written in terms of these length scales
as follows:
\begin{eqnarray}
R_{\hat{t}\hat{r}\hat{t}\hat{r}}
&=& \biggl(1- {b \over r} \biggr)\, \biggl[\pm { 1 \over {R_2 r_3} }
 + { 1 \over {R_2}^2 } \biggr] \, \pm {b \over r} \,
 \biggl(\pm  {1 \over {2 r_1 R_2} } - {1 \over {2 r R_2} } \biggr) \,,
                                     \label{eq:Rtrtr-ls}  \\
R_{\hat{t}\hat{\theta}\hat{t}\hat{\theta}}
&=& R_{\hat{t}\hat{\phi}\hat{t}\hat{\phi}}
= \pm \biggl(1- {b \over r} \biggr) \, {1 \over {r R_2} } \,,
                                            \label{eq:Rttheta-ls} \\
R_{\hat{r}\hat{\theta}\hat{r}\hat{\theta}}
&=& R_{\hat{r}\hat{\phi}\hat{r}\hat{\phi}}
= {b \over r } \, \biggl(\pm {1 \over {2 r r_1} } -
{1 \over {2 r^2} } \biggr) \,,   \label{eq:Rrtheta-ls}  \\
R_{\hat{\theta}\hat{\phi}\hat{\theta}\hat{\phi}}
&=& {b \over {r^3} } \,.             \label{eq:Rthetaphi-ls}
\end{eqnarray}
The choice of plus or minus signs in the various terms of the above
equations will depend on the signs of the derivatives of $b$ and $\Phi$,
which will in turn depend on the specific wormhole geometry.

Let the
magnitude of the maximum curvature component be $R_{max}$. Since the
largest value of $(1-b/r)$ and of $b/r$ is $1$, an examination
of Eqs.~(\ref{eq:Rtrtr-ls})-~(\ref{eq:Rthetaphi-ls})
shows that $R_{max} \lprox 1/({r_m}^2)$.
Therefore the smallest proper radius of curvature (which is also the
coordinate radius of curvature in an orthonormal frame) is:
\begin{equation}
r_c \approx {1 \over {\sqrt {R_{max}}} }  \gprox r_m \,.  \label{eq:l_c}
\end{equation}
Our length scales at the throat become:
\begin{equation}
\bar r_0 =r_0 \,; \,\, r_1 = \biggl| { r_0 \over {b'_0} } \biggr| \,; \,\,
R_2=  { {r_2} \over |{\Phi_0}| }  \,; \,\,
r_3= \biggl| { {\Phi'_0} \over {\Phi''_0} } \biggr| \,.  \label{eq:ls;r0}
\end{equation}
At the throat of the wormhole Eqs.~(\ref{eq:Rtrtr-ls})-
{}~(\ref{eq:Rthetaphi-ls}) simplify to
\begin{eqnarray}
R_{\hat{t}\hat{r}\hat{t}\hat{r}}|_{r_0}
&=& \pm {1 \over {2 r_0 R_2} } \pm {1 \over {2 r_1 R_2} }  \,,
                                     \label{eq:Rtrtr-ls;r0}  \\
R_{\hat{t}\hat{\theta}\hat{t}\hat{\theta}}|_{r_0}
&=& R_{\hat{t}\hat{\phi}\hat{t}\hat{\phi}}|_{r_0}
= 0  \,,                             \label{eq:Rttheta-ls;r0} \\
R_{\hat{r}\hat{\theta}\hat{r}\hat{\theta}}|_{r_0}
&=& R_{\hat{r}\hat{\phi}\hat{r}\hat{\phi}}|_{r_0}
= \pm {1 \over {2 r_0 r_1} } - {1 \over {2{r_0}^2} } \,,
                                    \label{eq:Rrtheta-ls;r0} \\
R_{\hat{\theta}\hat{\phi}\hat{\theta}\hat{\phi}}|_{r_0}
&=& {1 \over {r_0}^2 } \,.             \label{eq:Rthetaphi-ls;r0}
\end{eqnarray}
(At the throat, the length scale
$r_3$ does not explicitly appear in the curvature components.) Again,
we see that $R_{max} \lprox 1/({r_m}^2)$ and $r_c \gprox r_m$.

   We wish to work in a small spacetime volume around the throat of the
wormhole such that all dimensions of this volume are much smaller than $r_c$,
the smallest proper radius of curvature anywhere in the region. Thus, in the
absence of boundaries, spacetime can be considered to be
approximately Minkowskian in this region, and we should be able to
apply our QI-bound.

\section{Specific Examples}
\label{sec:SE}
  To develop physical intuition for the general case, as well as to get
a feeling for the magnitudes of the numbers involved, in this section
we apply our bound to a series of specific examples.

\subsection{$\Phi=0\,,b={r_0}^2/r$ Wormholes}
  This is a particularly simple wormhole which is discussed in
Box 2 and the bottom left-hand column of p. 400 of Ref. \cite{MT}.
In terms of the proper radial distance $l(r)$, the metric is:
\begin{equation}
ds^2= -dt^2 + dl^2
                + ({r_0}^2 +l^2) \,({d\theta}^2+
{\rm sin}^2\theta\,{d\phi}^2)\,,     \label{eq:b=b0^2/r}
\end{equation}
where $l=\pm(r^2-{r_0}^2)$. (Recall that $l=0$ at the throat.)
The stress-tensor components are given by
\begin{equation}
\rho= p_r= -P=\,- { {r_0}^2 \over {8\pi r^4} }=
- { {r_0}^2 \over {8\pi {({r_0}^2 + l^2)}^2 } } \,.   \label{eq:T;b0^2/r}
\end{equation}
The curvature components are
\begin{equation}
R_{\hat{\theta}\hat{\phi}\hat{\theta}\hat{\phi}}
=-R_{\hat{l}\hat{\theta}\hat{l}\hat{\theta}}
=-R_{\hat{l}\hat{\phi}\hat{l}\hat{\phi}}
={ {r_0}^2 \over { {({r_0}^2 + l^2)}^2 } } \,.   \label{eq:R;b0^2/r}
\end{equation}
Note that all the curvature components are equal in
magnitude, and have their maximum magnitude $1/({r_0}^2)$ at the throat.
The same holds true for the stress-tensor components. At the throat,
our length scales are $\bar r_0=r_0=r_1$, so $r_c=r_0$.

Let us apply our QI-bound to a static observer at $r=r_0$. (Recall that
such an observer is geodesic.) Since the energy density seen by this static
observer is constant we have
\begin{equation}
{{\tau_0} \over \pi}\, \int_{-\infty}^{\infty}\,
{{\langle T_{\mu\nu} u^{\mu} u^{\nu}\rangle\, d\tau}
\over {{\tau}^2+{\tau_0}^2}} = \rho_0 \gprox
-{c \over { {\tau_0}^4}} \,, \label{eq:QI;b0^2/r}
\end{equation}
where $c \equiv 3/(32 {\pi}^2)$, $\tau$ is the observer's proper time, and
$\tau_0$ is the sampling time. Choose our sampling time to be:
$\tau_0 = f r_m=f r_0 \ll r_c$, with $f \ll 1$. Substitution into
Eq.~(\ref{eq:QI;b0^2/r}) yields
\begin{equation}
r_0 \lprox { {l_p} \over {2 f^2} } \,, \label{eq:r_0;B1}
\end{equation}
where $l_p$ is the Planck length. Here it is fairly obvious that any
reasonable choice of $f$ gives a value of $r_0$ which is not much larger
than $l_p$. For example, for $f \approx 0.01$, $r_0 \lprox 10^4 \, l_p
= 10^{-31}  \,m$. Note from Eqs.~(\ref{eq:T;b0^2/r}) and
{}~(\ref{eq:R;b0^2/r}) that if we choose our spacetime region to be such that
$l \ll r_0$, then the curvature and stress-tensor components do not change
very much.

\subsection{$\Phi=0,\,b=r_0=const$ Wormholes}
For this wormhole $\Phi=0$ and $b=const$, so $b'=0$, and therefore
$\rho=0$. This is a special case of the ``zero density'' wormholes \cite{ZD}.
Here $g_{tt}$ is the same as for Minkowski
spacetime, while the spatial sections are the same as those of
Schwarzschild. The energy density and radial pressure seen by a static
observer are:
\begin{equation}
\rho=0\,;\,\, p_r= -{r_0 \over {8\pi r^3} }\,.  \label{T;bconst}
\end{equation}
Since the energy density is zero in the static frame, to obtain a bound
we boost to the frame of a radially moving geodesic observer. The energy
density in the boosted frame is, by a Lorentz transformation,
\begin{equation}
T_{\hat 0' \hat 0'} = \rho' = {\gamma^2}\,(\rho + v^2 \,p_r) \,,
                                                \label{eq:def-RHO'}
\end{equation}
where $v$ is the velocity of the boosted observer relative to the static
frame, and $\gamma=(1-v^2)^{-1/2}$. In our case, we have
\begin{equation}
\rho' = - { {\gamma^2 \, v^2 r_0 } \over {8\pi r^3} } \,.
                                                \label{eq:rho'_0;bconst}
\end{equation}
Note that in this case {\it any} non-zero $v$ gives $\rho'<0$, in contrast
to the discussion surrounding Eq. (57) of Ref. \cite{MT}.
The non-zero curvature components in the static frame are
\begin{equation}
R_{\hat{r}\hat{\theta}\hat{r}\hat{\theta}} =
R_{\hat{r}\hat{\phi}\hat{r}\hat{\phi}} = -{r_0 \over {2 r^3} } \,;
\,\, R_{\hat{\theta}\hat{\phi}\hat{\theta}\hat{\phi}} =
{r_0 \over {r^3} } \,.                          \label{Rcomp;bconst}
\end{equation}
Here the only relevant length scale is $r_0$, since
$\Phi=0$ and $b=r_0$ everywhere. For a general wormhole, when we boost
to the radially moving frame, we have
\begin{eqnarray}
R_{\hat{0'}\hat{1'}\hat{0'}\hat{1'}}
&=& R_{\hat{t}\hat{r}\hat{t}\hat{r}} \,,  \nonumber \\
R_{\hat{2'}\hat{0'}\hat{2'}\hat{0'}} &=&
R_{\hat{3'}\hat{0'}\hat{3'}\hat{0'}} \nonumber \\
  &=& { {\gamma}^2 \over {2 r^2} } \, \biggl[ v^2\,
\biggl( b'- {b \over r} \biggr) \,+\, 2(r-b)\,\Phi' \biggr] \,,
                                                 \nonumber \\
R_{\hat{2'}\hat{1'}\hat{2'}\hat{1'}} &=&
R_{\hat{3'}\hat{1'}\hat{3'}\hat{1'}} \nonumber \\
  &=& { {\gamma}^2 \over {2 r^2} } \, \biggl[
\biggl( b'- {b \over r} \biggr) \,+\, 2v^2\,(r-b)\,\Phi' \biggr] \,,
                                                      \nonumber \\
R_{\hat{2'}\hat{0'}\hat{2'}\hat{1'}} &=&
R_{\hat{3'}\hat{0'}\hat{3'}\hat{1'}} \nonumber \\
  &=& { {{\gamma}^2 v} \over {2 r^2} } \, \biggl[
\biggl( b'- {b \over r} \biggr) \,+\, 2(r-b)\,\Phi' \biggr] \,,
                                                      \nonumber \\
R_{\hat{2'}\hat{3'}\hat{2'}\hat{3'}} &=&
R_{\hat{\theta}\hat{\phi}\hat{\theta}\hat{\phi}}  \,.
                                               \label{eq:Rboost}
\end{eqnarray}
In the present case the non-zero components in the primed frame are
\begin{eqnarray}
R_{\hat{2'}\hat{0'}\hat{2'}\hat{0'}} &=&
R_{\hat{3'}\hat{0'}\hat{3'}\hat{0'}}
  = -{ {{\gamma}^2\,v^2\,r_0} \over {2 r^3} } \,,   \nonumber \\
R_{\hat{2'}\hat{1'}\hat{2'}\hat{1'}} &=&
R_{\hat{3'}\hat{1'}\hat{3'}\hat{1'}}
  = -{ {{\gamma}^2 \,r_0} \over {2 r^3} } \,,   \nonumber \\
R_{\hat{2'}\hat{0'}\hat{2'}\hat{1'}} &=&
R_{\hat{3'}\hat{0'}\hat{3'}\hat{1'}}
  = -{ {{\gamma}^2\, v  \,r_0} \over {2 r^3} } \,,   \nonumber \\
R_{\hat{2'}\hat{3'}\hat{2'}\hat{3'}} &=&
R_{\hat{\theta}\hat{\phi}\hat{\theta}\hat{\phi}}
= {r_0 \over {r^3} } \,.   \label{eq:Rboost;bconst}
\end{eqnarray}

   In the vicinity of the throat, the magnitude of the maximum curvature
component in the boosted frame is
$R'_{max} \lprox {\gamma}^2/({r_0}^2)$, and therefore the smallest local
proper radius of curvature in that frame is $r'_c \gprox r_0 / {\gamma}$.
Apply our QI-bound to the boosted observer and take
$\tau_0 = {fr_0 / {\gamma}} \ll r'_c$, for $f \ll 1$. Since the
energy density does not change much over this timescale, we may write
\begin{equation}
{{\tau_0} \over \pi}\, \int_{-\infty}^{\infty}\,
{{\langle T_{\mu\nu} u^{\mu} u^{\nu}\rangle\, d\tau}
\over {{\tau}^2+{\tau_0}^2}} \approx \rho'_0 \gprox
-{c \over { {\tau_0}^4}} \,, \label{eq:QI;bconst}
\end{equation}
which leads to
\begin{equation}
r_0 \lprox {\gamma \over {2 f^2 \, v}}\, l_p  \,.
                                     \label{eq:B1;bconst}
\end{equation}
In this case, any non-zero $v$ gives us a bound, but we can find
the optimum bound by minimizing ${\gamma}/v$, a procedure which
yields $v=1/{\gamma}= 1/{\sqrt{2}}$ and
\begin{equation}
r_0 \lprox { {l_p} \over {f^2} } \,. \label{eq:B2;bconst}
\end{equation}
Equation~(\ref{eq:B2;bconst}) is essentially the same as
Eq.~(\ref{eq:r_0;B1}), which was the bound we obtained in the
$\Phi=0, b={r_0}^2/r$ case.

\subsection{``Absurdly Benign'' Wormholes}

Classically, one can design a wormhole so that the exotic
matter is confined to an arbitrarily small region around the throat.
MT call this an ``absurdly benign''  wormhole. It is given by the
choices $\Phi = 0$ everywhere and
\begin{eqnarray}
b(r) &=& r_0[1-{(r-r_0)}/{a_0}]^2\,, \qquad
                  {\rm for}\,\, r_0\leq r\leq r_0+a_0 \,,   \nonumber \\
     &=& 0\,,  \qquad {\rm for} \,\, r\geq r_0+a_0 \,.    \label{eq:AB}
\end{eqnarray}
For $r_0 \leq r < r_0+a_0$,
\begin{eqnarray}
\rho &=& -\frac{r_0}{4\pi\, r^2\,a_0^2}\, (a_0+r_0-r) <0\,, \\
                                                     \label{eq:AB;rho}
p_r &=& -\frac{r_0}{8\pi\, r^3\,a_0^2}\, (a_0+r_0-r)^2 \,, \\ \label{eq:AB;p_r}
P &=& -{1\over2}(\rho+p_r)\,.                \label{eq:AB;P}
\end{eqnarray}
For $r\geq r_0+a_0$, the spacetime is Minkowski, and
$\rho=p_r=P=0$. The quantity $a_0$ represents the thickness in $r$
(on one side of the wormhole) of the negative energy region. Evaluation
of the curvature components using Eq.~(\ref{eq:AB}) shows that they
have maximum magnitude at the throat where
\begin{eqnarray}
R_{\hat{r}\hat{\theta}\hat{r}\hat{\theta}}|_{r_0}
&=& R_{\hat{r}\hat{\phi}\hat{r}\hat{\phi}}|_{r_0}
= -{1 \over {a_0 r_0} } - {1 \over {2{r_0}^2} } \,,
                                    \label{eq:Rrtheta;AB} \\
R_{\hat{\theta}\hat{\phi}\hat{\theta}\hat{\phi}}|_{r_0}
&=& {1 \over {r_0}^2 } \,.             \label{eq:Rthetaphi;AB}
\end{eqnarray}
At the throat, our length scales become
\begin{equation}
\bar r_0 = r_0 \,; \,\,r_1= {a_0 \over 2} \,,  \label{eq:AB;ls}
\end{equation}
and $r_m=min(r_0,r_1)$. Again we see that $R_{max} \lprox 1/({r_m}^2)$,
and so the smallest local radius of curvature is $r_c \gprox r_m$.

Application of our QI-bound to a static observer at the throat yields
\begin{equation}
\rho_0 = - {1 \over {4\pi a_0 r_0} } \gprox - {c \over {{\tau_0}^4} }\,.
                                                \label{eq:AB;QI}
\end{equation}
Although this  wormhole was designed for maximum confinement of the
negative energy near the throat, i.e., $a_0 \ll r_0$, there is
nothing in principle to keep us from choosing $a_0 \gprox r_0$. In what
follows, we shall consider both situations. First assume $a_0<r_0$.
We then choose our sampling time to be: $\tau_0 = f a_0$, where $f \ll 1$.
Equation~(\ref{eq:AB;QI}) then yields
\begin{equation}
a_0 \lprox {\biggl( {{ r_0} \over {8 f^4\, l_p} } \biggr) }^{1/3} \,l_p\,.
                                                \label{eq:AB;Ba0}
\end{equation}
A reasonable choice of $f$ is $f \approx 0.01$. For a small
``human sized'' wormhole with $r_0 \approx 1\,m$, our bound gives
$a_0 \lprox 10^{14}\,l_p \approx 10^{-21}\,m \approx 10^{-6}$ fermi,
or approximately a millionth of the proton radius. The situation does
not improve much for larger wormholes. For $r_0 \approx 1 $ light year,
$a_0 \lprox 2 \times 10^{19} \, l_p \approx 0.2$ fermi. With
$r_0 \approx 10^{5}$ light years, $a_0 \lprox 10^{21}\, l_p
\approx 10^{-14}\,m$. So even with a throat radius the size of a
galaxy, the negative energy must be distributed in a band no thicker
than about $10$ proton radii. Now suppose that $r_0 < a_0/2$, so
that $r_m=r_0$. In that case, we choose $\tau_0 =f r_0$, and
our bound gives
\begin{equation}
r_0 \lprox {\biggl( {{ a_0} \over {8 f^4\, l_p} } \biggr) }^{1/3} \,l_p\,,
                                                \label{eq:AB;Bb0}
\end{equation}
i.e., $a_0$ and $r_0$ are simply interchanged. Therefore, the
same numerical examples just discussed now apply to $r_0$, for given
choices of $a_0$. For example, when $a_0 \approx 1$ light year, now the
throat size is less than about $0.2$ fermi, so that even for very large $a_0$,
$r_0$ must be extremely small. When $a_0 \approx r_0$, the bound on $r_0$
is essentially Eq.~(\ref{eq:r_0;B1}). One might worry that since $a_0$ is the
coordinate thickness in $r$ of the negative energy density, it might
not be a good measure of the proper radial thickness of the negative
energy density band seen by the static observer. In fact, a detailed
calculation shows that $a_0$ {\it is} the proper thickness in this case,
to within factors of order unity.

\subsection{``Proximal Schwarzschild'' Wormholes}
 Another special case of a zero density wormhole is the ``proximal
Schwarzschild'' wormhole \cite{PS}. Here $b=r_0=const$, and
$g_{tt}$ is only slightly different from that of Schwarzschild.
The metric in this case is:
\begin{equation}
 ds^2=- \biggl( 1- {r_0 \over r} + {\epsilon \over r^2}
\biggr)\,dt^2+{{dr^2}\over{(1- {r_0 / r}) } }
           +r^2({d\theta}^2+ {\rm sin}^2\theta\,{d\phi}^2)\,.
                                                \label{eq:PSM}
\end{equation}
We recover the Schwarzschild solution for $\epsilon=0$; however, any
$\epsilon>0$ gives us a wormhole. The energy density and radial pressure
seen by a static observer are
\begin{eqnarray}
\rho &=& 0\,,      \\   \label{eq:PS;rho}
p_r &=& -{ \epsilon \over {8\pi r^4} }
    { {(2 - {r_0}/{r})} \over {( 1- {r_0 / r}  +
{\epsilon / r^2})  } }  \,.  \label{eq:PS;p_r}
\end{eqnarray}
We will assume that $\sqrt{\epsilon} \ll r_0$, hence the radial pressure
is highly peaked near the throat. Here the proper distance from $r=r_0$ to
$r=r_0 + \sqrt{\epsilon}$, corresponding to the coordinate thickness
$\sqrt{\epsilon}$, is
\begin{equation}
\Delta l = \int_{r_0}^{r_0 + \sqrt{\epsilon}} \,
{ {dr} \over {\sqrt{1- {r_0 / r}} } } \approx
\int_{r_0}^{r_0 + \sqrt{\epsilon}} \,
{ {\sqrt{r_0} \,dr} \over {\sqrt{r- r_0} } } \,=
2\, \sqrt{r_0 \sqrt{\epsilon} }   \,.            \label{Deltal;PS}
\end{equation}
A disadvantage of this wormhole is that it entails extremely large
redshifts. The metric coefficient $g_{tt}$ is very close to that of
Schwarzschild, and therefore this wormhole is very close to having
a horizon at its throat.

In the region
$r_0 \leq r \leq r_0 + \sqrt{\epsilon}$, the curvature components have
their maximum magnitudes at the throat (except
$R_{\hat{t}\hat{\theta}\hat{t}\hat{\theta}} =
R_{\hat{t}\hat{\phi}\hat{t}\hat{\phi}}$, which vanish there). At $r=r_0$,
\begin{eqnarray}
R_{\hat{t}\hat{r}\hat{t}\hat{r}}|_{r_0}
&\approx& {1 \over {4\epsilon} }   \,,
                                     \label{eq:Rtrtr-PS;r0}  \\
R_{\hat{r}\hat{\theta}\hat{r}\hat{\theta}}|_{r_0}
&=& R_{\hat{r}\hat{\phi}\hat{r}\hat{\phi}}|_{r_0}
= - {1 \over {2{r_0}^2} } \,,
                                    \label{eq:Rrtheta-PS;r0} \\
R_{\hat{\theta}\hat{\phi}\hat{\theta}\hat{\phi}}|_{r_0}
&=& {1 \over {r_0}^2 } \,.             \label{eq:Rthetaphi-PS;r0}
\end{eqnarray}
Our length scales at the throat become
\begin{equation}
\bar r_0=r_0\,;\,\,r_1=\infty\,;\,\,R_2\approx { {2\epsilon} \over r_0}\,;
\,\,r_3 \approx { {\epsilon} \over {r_0} }\,.    \label{PS;ls}
\end{equation}
Note that $r_1=\infty$ is due to the fact that $b'=0$. Although we
can write the curvature components in terms of these length scales,
in this case the smallest radius of curvature in the
static frame is $r_c \approx 1/{\sqrt{R_{max}}}
\approx 2 \sqrt{\epsilon}$,
which is larger than our smallest length scale $r_m=r_3$.
Thus we will get a stronger bound if we frame our argument in terms
of $r_c \approx 2 \sqrt{\epsilon}$.

Since the energy density is zero in the static frame, we must apply
our bound in the frame of a boosted observer passing through the
throat. The curvature tensor components in this frame are
\begin{eqnarray}
R_{\hat{1'}\hat{0'}\hat{1'}\hat{0'}}|_{r_0} &=&
R_{\hat{t}\hat{r}\hat{t}\hat{r}}|_{r_0}
\approx {1 \over {4\epsilon} }   \,,  \nonumber \\
R_{\hat{2'}\hat{0'}\hat{2'}\hat{0'}}|_{r_0} &=&
R_{\hat{3'}\hat{0'}\hat{3'}\hat{0'}}|_{r_0}
  =- { {{\gamma}^2\,v^2} \over {2 {r_0}^2} } \,,   \nonumber \\
R_{\hat{2'}\hat{1'}\hat{2'}\hat{1'}}|_{r_0} &=&
R_{\hat{3'}\hat{1'}\hat{3'}\hat{1'}}|_{r_0}
  = -{ {{\gamma}^2} \over {2 {r_0}^2} } \,,   \nonumber \\
R_{\hat{2'}\hat{0'}\hat{2'}\hat{1'}}|_{r_0} &=&
R_{\hat{3'}\hat{0'}\hat{3'}\hat{1'}}|_{r_0}
  =- { {{\gamma}^2\,v} \over {2 {r_0}^2} } \,,   \nonumber \\
R_{\hat{2'}\hat{3'}\hat{2'}\hat{3'}}|_{r_0} &=&
R_{\hat{\theta}\hat{\phi}\hat{\theta}\hat{\phi}}|_{r_0}
= {1 \over {{r_0}^2} } \,.   \label{eq:Rboost;PS}
\end{eqnarray}
Which of these components has the maximum magnitude depends on
whether $\sqrt{2\epsilon}$ is greater than or less than $r_0/{\gamma}$.

First consider the case: $\sqrt{2\epsilon}<r_0/{\gamma}$. Then
$R'_{max} = R_{max} \approx 1/({4\epsilon})$, and
$r'_c = r_c \approx 2 \sqrt{\epsilon}$. Take
$\tau_0= f r'_c \approx 2f \sqrt{\epsilon}$,
with $f<<1$. The energy density in the boosted frame should be
approximately constant over this sampling time.
Therefore, our QI-bound gives
\begin{equation}
\rho'_0=- { {{\gamma}^2 \,v^2} \over {8\pi {r_0}^2 } } \,
\gprox - {c \over {{\tau_0}}^4 } \,,      \label{eq:QI;rho'}
\end{equation}
and hence
\begin{equation}
{ {\sqrt\epsilon} \over {r_0} } \lprox
{\biggl( { {1-v^2} \over {64\,v^2\,f^4} } \biggr)}^{1/4} \,
{\biggl( { l_p \over r_0} \biggr) }^{1/2} \,.    \label{eq:QI1;PS}
\end{equation}
By making $v$ arbitrarily close to $1$, we can make the right-hand side
of the bound arbitrarily small.

It may be more appropriate to express the width of the band of exotic
matter in the static frame in terms of proper length, rather than
coordinate length. Using Eq.~(\ref{Deltal;PS}), our bound
Eq.~(\ref{eq:QI1;PS}) can be rewritten as
\begin{equation}
{ {\Delta l} \over {r_0} } \lprox
{\biggl( { {1-v^2} \over {v^2\,f^4} } \biggr)}^{1/8} \,
{\biggl( { l_p \over r_0} \biggr) }^{1/4} \,.    \label{eq:QI1-l;PS}
\end{equation}
In this form, the bound is quite a bit weaker, due to the smaller
powers on the right-hand side of the inequality. We can still
in principle make the right-hand side arbitrarily small,
albeit only by choosing $v$ {\it exceedingly} close to $1$. However, our
bound must hold for {\it any} boosted observer. Consequently, for the case
$\sqrt{2\epsilon} < r_0/{\gamma}$, proximal Schwarzschild wormholes
with any finite value of $\Delta l$ would seem to
be physically excluded.

 Next consider the case where $\sqrt{2\epsilon}>r_0/{\gamma}$. Then
$R'_{max} = {\gamma}^2 / (2 {r_0}^2) $, and the
smallest local radius of curvature in the boosted frame is
$r'_c \approx {\sqrt{2}\,r_0} / {\gamma}$. Take $\tau_0=f r'_c$,
with $f \ll 1$. Application of our bound in this case yields
\begin{equation}
r_0 \lprox { {\sqrt{2 \pi c} } \over f^2 } \,
\biggl({ {\gamma} \over v} \biggr) \,.
\end{equation}
We get the optimum bound by minimizing ${ {\gamma} / v}$, i.e.,
\begin{equation}
r_0 \lprox { {l_p} \over {2 f^2} } \,, \label{eq:QI2;PS}
\end{equation}
which is the same as the bound we obtained in the
$\Phi=0, b={r_0}^2/r$ case.

\subsection{The Morris-Thorne-Yurtsever Wormhole}

 Morris, Thorne, and Yurtsever (MTY) \cite{MTY} have discussed a wormhole
consisting of an $r_0 = Q= M$ Reissner-Nordstr\"om (RN) metric with
a pair of spherical charged Casimir plates positioned on each side of
the throat within a very small proper distance, $s$, of one another.
That is, the spacetime is extreme RN from each plate out to
$r=\infty$, and approximately flat between the plates. The
Casimir energy density between the plates is negative, while the
stress-energy of the external classical electromagnetic field
is ``near exotic'', i.e., $(\rho_c + p_c)|_{EM} =0$. For
$r \geq r_0 + \delta$, the metric has the extreme RN form
\begin{equation}
 ds^2=- {\biggl( 1- {M \over r}\biggr)}^2\,dt^2
+{{dr^2}\over{ {(1- {M / r})}^2 } }
           +r^2({d\theta}^2+ {\rm sin}^2\theta\,{d\phi}^2)\,.
                                                \label{eq:ERNM}
\end{equation}

MTY show that for this wormhole
\begin{equation}
p_0 = -\frac{1}{8\pi r_0^2} = p_{Casimir} \,.
\end{equation}
Then because $\rho_{Casimir} = \frac{1}{3} p_{Casimir}$ and $\rho_{Casimir}
= \rho_0$, it follows that $\rho_0 = -(24 \pi r_0^2)^{-1}$.
If we apply our bound to a (very tiny) static
observer at the throat, we obtain
\begin{equation}
-\frac{1}{24\pi {r_0}^2}
\gprox -\frac{c}{ {\tau_0}^4 }\,.          \label{eq:QI1-MTY}
\end{equation}
Because spacetime is approximately flat between the plates, the constraint
on the choice of $\tau_0$ is that discussed in the Casimir effect example in
Sec.~\ref{sec:QI}. Take $\tau_0 = f s$, whereupon we find:
\begin{equation}
r_0 \gprox f^2 \, s^2  \,.           \label{eq:QIB;MTY}
\end{equation}
A reasonable choice of $f$ in this case would seem to be
$f \approx 0.1$. For $s \approx 10^{-10} \,cm
\approx 10^{23} \,l_p$, one finds that: $r_0 \gprox 10^{44}\, l_p
\sim 0.01 \, A.U.$. Since MTY calculate $r_0 \approx 1\, A.U.$ for
a plate separation of $s \approx 10^{-10} \,cm$, this wormhole
satisfies our bound.

However, this wormhole has a number of undesirable features. First,
in order to traverse it, an observer must go through the plates.
This implies that ``holes'' or ``trapdoors'' must be cut in the
plates to allow passage. Second, because the plates are located at
$r \approx r_0 + \delta$, with $\delta \approx 10^{-10} \,cm$ (neglecting
the thickness of the plates), this
wormhole is extremely close to being a black hole, i.e.,
$|g_{tt}|_{r_0 + \delta} = {(1-M/r)}^2 \approx {\delta}^2/M^2$.
Infalling photons with frequency-at-infinity $\omega_{\infty}$
will have local frequency $\omega_{local} \approx
\omega_{\infty} \, (M/ \delta)$, as measured by a static observer on the
plate. For $\delta \approx 10^{-10} \,cm$ and $r_0 \approx M
\approx 1 \, A.U. \approx 10^{13} \,cm$, we have: $\omega_{local} \approx
10^{23}\, \omega_{\infty}$. A typical infalling $3K$ photon in the
cosmic microwave background radiation, upon arriving at one of
the plates, would get blueshifted to a temperature $T_{local}
\approx 10^{23}\,K$. A $0.1 \,MeV$ $\gamma$-ray photon would get
blueshifted to $E_{local} \approx 10^{19} \,GeV \approx E_p$, where
$E_p$ is the Planck energy.
Stray cosmic ray particles, with typical energies of
$\sim 1\, GeV$, falling into the wormhole will have $E_{local}
\approx E_{\infty} \,(M/ \delta) \approx 10^{23} \,GeV
\approx 10^4 \,E_p$. A static observer just outside the plates
would likely be incinerated by infalling radiation. Similarly,
the plates would have to be constructed out of material capable
of withstanding these large energies. In addition, the large
local impacts of infalling radiation and particles on the plates
will tend to push them together, thereby upsetting the force balance,
and hence will probably destabilize the wormhole. One {\it could}
imagine elaborate radiation shielding constructed around a large
region far away from, but enclosing, the wormhole. However, if the
wormhole is unstable to infalling radiation, then infalling spaceships
would seem like an even more remote possibility.

\section{General Bounds for Wormholes}
\label{sec:GB}

 In this section QI-bounds will be formulated on the relative
size scales of arbitrary static, spherically symmetric, MT wormholes,
i.e., no assumptions
will be made about the specific forms of $\Phi(r)$ and $b(r)$. We work
in the vicinity of the throat and analyze two general subcases:
1) $b'_0<0$, and 2) $b'_0 \geq 0$. Let $r_m$ be the smallest of the
length scales: $\bar r_0,r_1,R_2,r_3$, in this region. We saw
from Eqs.~(\ref{eq:Rtrtr-ls})-
{}~(\ref{eq:Rthetaphi-ls}) in Sec.~\ref{sec:REVMT} that
the magnitude of the maximum curvature component in this region
is $R_{max} \lprox 1/({r_m}^2)$. It follows that the smallest proper
radius of curvature (in the static orthonormal
frame) is $r_c \approx {1 / {\sqrt {R_{max}}} }  \gprox r_m$.
Spacetime can be considered to be approximately flat in this
region, and therefore our QI-bound should be applicable.

\subsection{Case 1): $b'_0<0$}
 Since $b'_0<0$, the energy density is negative at the
throat, so we can apply our bound to a static observer at the throat.
This observer is geodesic, and the energy density is constant, so we have
\begin{eqnarray}
{{\tau_0} \over \pi}\, \int_{-\infty}^{\infty}\,
{{\langle T_{\mu\nu} u^{\mu} u^{\nu}\rangle\, d\tau}
\over {{\tau}^2+{\tau_0}^2}} &=& \rho_0    \nonumber  \\
&=& { {b'_0} \over {8\pi {r_0}^2} }  \gprox
-{c \over { {\tau_0}^4}} \,, \label{eq:QI1;b'0<0}
\end{eqnarray}
where again $c \equiv 3/(32 {\pi}^2)$, $\tau$ is the observer's proper
time, and $\tau_0$ is the sampling time. Choose our sampling time to be:
$\tau_0=f r_m \ll  r_c$, with $f \ll 1$. Our QI then becomes
\begin{equation}
r_m \lprox  {\biggl( \frac{8\pi c}{|b'_0|} \biggr)}^{1/4} \,
\frac{\sqrt{r_0}}{f} \,.
\end{equation}
Since our observer is static at the throat, we may write $|b'_0| = r_0/ r_1$,
and use $8\pi c \approx 1/4$, to get
\begin{equation}
r_m \lprox \frac{{(r_0 r_1)}^{1/4}}{f} \,,        \label{eq:GQI1;b'0<0}
\end{equation}
or alternatively,
\begin{equation}
{r_m \over r_0} \lprox {1 \over f} {\biggl(
\frac{r_1\,{l_p}^2}{{r_0}^3} \biggr)}^{1/4}   \,.  \label{eq:GQ2;b'0<0}
\end{equation}

Now examine specific cases. For $r_m=r_0$, we have: $r_0/r_1
\lprox f^{-4/3} \,{(l_p/r_1)}^{2/3}$. As an example, if $r_1
\approx 1\,m$ and for $f \approx 0.01$, then $r_0/r_1
\lprox 10^{-21}$.
Even if we choose $f$ to be very small this large discrepancy in
length scales will not change much, and only increases as
$r_1$ increases. For $r_m=r_1$, $r_1/r_0 \lprox
f^{-4/3}\,{(l_p/r_0)}^{2/3}$, so for $r_0
\approx 1\,m$ and $f \approx 0.01$, $r_1/r_0 \lprox 10^{-21}$.
Again the problem only gets worse as the throat size, $r_0$, increases.
When $r_m=r_0=r_1$: $r_0 \lprox l_p/f^2$; for $f \approx 0.01$,
$r_0 \lprox 10^4\,l_p \sim 10^{-31} \,m$. For $\Phi = const$ wormholes,
the only relevant length scales are $r_0$ and $r_1$. The above results imply
that when $b'_0 < 0$ these wormholes are extremely unlikely, unless one
is willing to accept a huge discrepancy in length scales. For $r_m=R_2$ and
$R_2 \leq r_0 \leq r_1$, from Eq.~(\ref{eq:GQI1;b'0<0}) we have that:
$R_2 /r_1 \lprox (1/f)\,{(r_0 \,{l_p}^2 /{r_1}^3)}^{1/4} \lprox
(1/f)\,{(l_p/r_1)}^{1/2}$. For $r_m=R_2$ and $R_2 \leq r_1 \leq r_0$:
$R_2 /r_0 \lprox (1/f)\,{(l_p/r_0)}^{1/2}$. An identical argument yields
similar inequalities for the case where $r_m=r_3$. Thus we find that if
$r_0$ and/or $r_1$ are macroscopic, then the ratio of the minimum length
scale to the macroscopic length scale must be very tiny.

If the minimum
scale happens to be $R_2$, then our bounds imply that either $r_2$,
the scale over which $\Phi$ changes, is very small or $|\Phi|$ is very
large, or both. A situation in which $|\Phi|$ is very large near the
throat, assuming $\Phi(\infty) = 0$,
would not seem to be a desirable characteristic of a traversable
wormhole, as it implies very large redshifts or blueshifts
for a static observer at the throat. A large negative $\Phi_0$
implies that the spacetime is close to having a horizon at the throat.
A large positive $\Phi_0$ implies that photons of moderate
frequency fired outward by an observer at the throat would be
blueshifted to very high frequencies upon reaching distant observers.
In the latter case, observers must be shot inward with initially large
kinetic energies in order to reach the throat.

If some of the wormhole parameters change over
very short length scales, then it would seem from the ``tidal force
constraints'' (see Eqs. (49) and (50) of MT) that tidal accelerations
might also change over very short length scales. As a result, an
observer travelling through the wormhole could encounter potentially
wrenching tidal forces rather abruptly. None of these scenarios seem
terribly convenient for wormhole engineering.

\subsection{Case 2): $b'_0 \geq 0$}
  When $b'_0 \geq 0$, the energy density at the throat is non-negative
for static observers. To obtain a bound in this case, we Lorentz-transform
to the frame of a radially moving boosted observer at the throat. Since the
maximum magnitude curvature component in the static frame is
$R_{max} \lprox 1/({r_m}^2)$, in the boosted frame the curvature
component with the largest magnitude, $R'_{max}$, can be {\it no larger
than} about ${\gamma}^2/{r_m}^2$. Therefore, the smallest proper radius of
curvature in the boosted frame is $r'_c \approx 1/ \sqrt{R'_{max}}
\gprox r_m /{\gamma}$. Spacetime should be approximately flat in the
boosted frame on scales much less than $r'_c$. Hence let us take our
sampling time to be: $\tau_0 = f r_m /{\gamma} \ll  r'_c$, with
$f \ll 1$. The energy density in this frame should not
change much over the sampling time, so the application of our bound gives
\begin{eqnarray}
{{\tau_0} \over \pi}\, \int_{-\infty}^{\infty}\,
{{\langle T_{\mu\nu} u^{\mu} u^{\nu}\rangle\, d\tau}
\over {{\tau}^2+{\tau_0}^2}} & \approx &
\langle T_{\mu\nu} u^{\mu} u^{\nu}\rangle     \nonumber  \\
&=& \rho'_0 \gprox -{c \over { {\tau_0}^4}} \,. \label{eq:QI1;b'0 geq 0}
\end{eqnarray}
At the throat, from Eq.~(\ref{eq:def-RHO'}), the energy density in the
boosted frame is
\begin{equation}
\rho'_0 = {\gamma}^2 \, (\rho_0 + v^2\,p_0)
        =  { {\gamma}^2 \over {8\pi {r_0}^2} } \, (b'_0-v^2) \,.
                                            \label{eq:rho';GB}
\end{equation}
In order for $\rho'_0 < 0$, we must require $v^2 > b'_0$.
After making the required substitutions, we obtain
\begin{equation}
{r_m \over r_0} \lprox {\biggl(\frac{1}{v^2-b'_0}\biggr)}^{1/4} \,
\frac{\sqrt{\gamma}}{f}\,{\biggl(\frac{l_p}{r_0}\biggr)}^{1/2} \,.
                                     \label{eq:GB;b'0 geq 0}
\end{equation}

For $b'_0 \geq 0,\, 0 \leq b'_0 \leq 1$, since $b'_0 \leq 1$, which
follows from the fact that at the throat we must have
$\rho_0 + p_0 \leq 0$ \cite{VT}.
The quantity $b'_0$ is fixed by the wormhole geometry, whereas our
choice of $v^2$ is arbitrary, subject to $b'_0 <v^2<1$. Our
bound, Eq.~(\ref{eq:GB;b'0 geq 0}), is weakest when
$b'_0$ is extremely close to $1$. However,
this would seem to be a highly special case, $b'_0 =1$ corresponding
to the maximum possible positive energy density at the throat and to
$\rho_0 + p_0 = 0$, i.e., the null energy condition
$T_{\mu \nu} K^{\mu} K^{\nu} \geq 0$, applied to radial null
vectors is barely satisfied at the throat. The latter implies
that such a wormhole ``flares outward'' very slowly from the
throat (see, for example, Eq. (56) of MT). To see how close $b'_0$
must be to $1$ in order to significantly affect our bound,
a numerical example is instructive. Let
$b'_0 = 1 - 10^{-8}$, $v^2 = 1 - 10^{-9}$, and $f \approx 0.01$. For
$r_0 \approx 1 m \approx 10^{35}\, l_p$, we find that $r_m \lprox 10^{-11} m$.
Even for $r_0 \approx 1 A.U. \approx 10^{46}\, l_p$, we obtain
$r_m \lprox 10^{-6} m$. If we consider a more ``typical'' $b'_0 \geq 0$
to be in about the middle of the allowed range, say $b'_0 \approx 1/2$,
then if we choose $v^2 \approx 3/4$, it follows that
${(v^2-b'_0)}^{-1/4} \approx 1$. If we choose
$f \approx 0.01$, then we find: $r_m/r_0 \lprox 100 \,{(l_p/r_0)}^{1/2}$.
Even a much smaller choice of $f$ does not avoid the large discrepancy
in wormhole length scales.

The bound Eq.~(\ref{eq:GB;b'0 geq 0}) is a ``safe''
bound, but in specific cases it may not be the optimal bound, due
to our rather conservative condition on $R'_{max}$, i.e.,
$R'_{max} \lprox {\gamma}^2/{r_m}^2$, and hence on $\tau_0$. For example,
in cases where $R'_{max}=|R_{\hat{1'}\hat{0'}\hat{1'}\hat{0'}}|=
|R_{\hat{t}\hat{r}\hat{t}\hat{r}}|$,
such as the proximal Schwarzschild wormhole,
we can get a stronger bound than that obtained from the inequality for the
general case.

\section{Conclusions}
\label{sec:con}

 In this paper, we used a bound on negative energy density derived
in four-dimensional Minkowski spacetime to constrain static, spherically
symmetric traversable wormhole geometries. In Sec.~\ref{sec:QI},
we argued that the bound should also be applicable in curved spacetime on
scales which are much smaller than the minimum local radius of curvature
and/or the distance to any boundaries in the spacetime.
The upshot of our analysis is that either a wormhole must have a throat
size which is only slightly larger than the Planck length $l_p$, or there
must be large discrepancies in the length scales which characterize the
geometry of the wormhole. These discrepancies are typically of
order $(l_p/r_0)^{n}$, where $r_0$ is the throat radius and $n \lprox 1$.
They imply that generically the exotic matter is confined to an extremely
thin band, and/or that the wormhole geometry involves large redshifts
(or blueshifts). The first feature would seem to be rather physically
unnatural. Furthermore, wormholes in which the characteristics of the
geometry change over short length scales and/or entail large redshifts
would seem to present severe difficulties for traversability, such as large
tidal forces.

There are a number of possible ways to circumvent our conclusions. The primary
contributions to the exotic matter might come from the state-independent
geometrical terms of $\langle T_{\mu \nu} \rangle$. However, as discussed in
Sec.~\ref{sec:QI}, in this case the dimensionless coefficients of these terms
would have to be enormous to generate a wormhole of macroscopic size. One
possibility would be a model in which the effective values of these
coefficients are
governed by a new field $\phi$ in such a way that they are large only when
$\phi$ is large. It may then be possible to find self-consistent solutions in
which $\phi$ is large only in a very small region, and hence it is conceivable
that one might be able to create thin bands of negative energy in this way
\cite{Frolov}.
Our bound was strictly derived only for a massless, minimally coupled
scalar field, but we argued that similar bounds are likely to hold for
other massless and massive quantum fields. Another possible circumvention
of the bound might be to superpose the effects of many fields, each of
which satisfies the bound \cite{MP}.
For example, suppose we postulate $N$ fields,
each of which contribute approximately the same amount to our bound.
Then the right-hand side of the inequality Eq.~(\ref{eq:QI}) would be
replaced by $-N c/ {\tau_0}^4$. However, in practice $N$ has to be extremely
large in order to have a significant effect.
For example, in the case of the $\Phi=0,\,b=const$
wormholes discussed in Sec.~\ref{sec:SE}, the constraint on the throat size
becomes $r_0 \lprox \sqrt N / (2 f^2)$. For $f \approx 0.01$, $r_0 \lprox
\sqrt N\, 10^4\, l_p \sim \sqrt N\, 10^{-31} \,m$. Therefore, to get $r_0
\approx 1\,m$, we would need either $10^{62}$ fields or a few
fields for which the constant $c$ is many orders of magnitude larger than
$3/(32\, \pi^2)$. Neither of these possibilities seem very likely.
Lastly, it may be that the semiclassical theory breaks down above the Planck
scale, due to large stress-tensor fluctuations when the mean energy density
is negative \cite{Kuo,KF}.
In that case, it becomes difficult to predict what happens. However,
one might expect the timescale of such fluctuations to be of the order
of the minimum radius of curvature. Since our sampling time is chosen to be
much smaller than this, it may be that our analysis is unaffected by the
fluctuations.

We showed that the Morris-Thorne-Yurtsever \cite{MTY} wormhole was
compatible with our
bound. When this model was proposed some years ago, it was hoped that one
might eventually be able to do better at spreading the exotic matter out over
macroscopic dimensions. Our results indicate that this kind of wormhole might
be the generically allowed case. However as we pointed out, this wormhole
has undesirable features, such as large redshifts near the throat which may
pose problems for stability and traversability. It might seem that our
conclusions imply that the most physically reasonable wormholes are the
``thin-shell'' type \cite{VISSER-TS}. However, these models
are constructed by ``cutting and pasting'' two copies of (for example)
Minkowski or Schwarzschild spacetime, with a resulting $\delta$-function
layer of negative energy at the throat. (Note that in these wormholes, by
construction, the throat is not located at $b=r_0$.) Physically one does
not really expect infinitely thin layers of energy density and curvature
in nature \cite{VISSER-CW}. Such approximations are meant to be
idealizations of situations
in which the thickness of these layers are small compared to other relevant
length scales. Our results can be construed as placing upper bounds on
the actual allowed thicknesses of such layers of negative energy density.
We conclude that, unless one is willing to accept fantastically large
discrepancies in the length scales which characterize wormhole geometries,
it seems unlikely that quantum field theory allows macroscopic static
traversable wormholes \cite{MITCH}.
\vskip 1cm
\centerline{\bf Acknowledgements}
The authors would like to thank Arvind Borde, Eanna Flanagan, John
Friedman, Valery Frolov, Bernard Kay, Mike Morris, Niall O'Murchadha,
Adrian Ottewill, Matt Visser, and Bob Wald for helpful discussions.
TAR would like to thank the
members of the Tufts Institute of Cosmology for their kind hospitality
and encouragement while this work was being done.
This research was supported in part by NSF Grant No. PHY-9208805 (Tufts), by
the U.S. Department of Energy (D.O.E.) under cooperative agreement
\# DF-FC02-94ER40818 (MIT), and by a CCSU/AAUP Faculty Research Grant.


\begin{thebibliography}{--}
\bibitem{LOG} Note that the logarithmic dependence upon the curvature
causes the effective values of $c_1$ and $c_2$ to be running coupling
constants: For example, let $c_{1\,eff} = c_1 \, \log(R\,{\mu}^{-2})$.
Then $|c_{1\,eff}|$ increases slowly as $R$ increases. Between the Planck
scale and some larger length scale, $|c_{1\,eff}|$ increases by a factor
of $\log(l^2/{l_p}^2)$, if we take $\mu={l_p}^{-1}$. Thus, in our example
of a wormhole solution with $l \approx 1m$, if we wish to have
$|c_{1\,eff}| \approx 10^{70}$, then at the Planck scale $|c_1| \approx
10^{68}$, which is a marginal change.
\bibitem{PS} See Ref.\cite{VBOOK}, Sec. 13.4.3. Note that the right-hand side
of Eq. (13.88) of that reference is missing a factor of $(2-r_0/r)$ in the
numerator.
\bibitem{Frolov} This possibility was suggested to us by Valery Frolov.
\end{thebibliography}
\end{document}